\documentclass[fleqn,usenatbib]{mnras}

\usepackage{newtxtext,newtxmath}
\usepackage[T1]{fontenc}

\DeclareRobustCommand{\VAN}[3]{#2}
\let\VANthebibliography\thebibliography
\def\thebibliography{\DeclareRobustCommand{\VAN}[3]{##3}\VANthebibliography}

\usepackage{graphicx}	% Including figure files
\usepackage{amsmath}	% Advanced maths commands
\usepackage{newtxtext,newtxmath}

\newcommand{\LL}{\mathcal{L}}

\usepackage[dvipsnames]{xcolor}
\definecolor{mygreen}{rgb}{0.1, 0.75, 0.3}
\definecolor{myorange}{rgb}{0.7, 0.2, 0.}

\title[Modelling RSD with Lagrangian bias and simulations]{Modelling galaxy clustering in redshift space with a Lagrangian bias formalism and $N$-body simulations.}

\author[Pellejero Iba\~nez et al.]{
Marcos Pellejero Iba\~nez,$^{1}$\thanks{E-mail: mpellejero@dipc.org}
Jens St\"ucker,$^{1}$ 
Raul E. Angulo,$^{1,2}$ 
Matteo Zennaro,$^{1}$ 
Sergio Contreras$^{1}$ \newauthor
and Giovanni Aric\`o.$^{1,3}$
\\
$^{1}$Donostia International Physics Center (DIPC), Paseo Manuel de Lardizabal, 4, 20018 Donostia-San Sebasti\'an, Spain.\\
$^{2}$IKERBASQUE, Basque Foundation for Science, 48013, Bilbao, Spain.\\
$^{3}$Departamento de F\'isica Te\'orica, Universidad de Zaragoza, Pedro Cerbuna 12, 50009 Zaragoza, Spain.
}

\date{Accepted XXX. Received YYY; in original form ZZZ}

\pubyear{2015}

\begin{document}
\label{firstpage}
\pagerange{\pageref{firstpage}--\pageref{lastpage}}
\maketitle

\begin{abstract}
Improving the theoretical description of galaxy clustering on small scales is an important challenge in cosmology, as it can considerably increase the scientific return of forthcoming galaxy surveys -- e.g. tightening the bounds on neutrino masses and deviations from general relativity.
In this paper we propose and test a new model for the clustering of galaxies that is able to accurately describe redshift-space distortions even down to small scales. This model corresponds to a second-order perturbative Lagrangian bias expansion which is advected to Eulerian space employing a displacement field extracted from $N$-body simulations. Eulerian coordinates are then transformed into redshift space by directly employing simulated velocity fields augmented with nuisance parameters capturing various possible satellite fractions and intra-halo small-scale velocities. We quantify the accuracy of our approach against samples of physically-motivated mock galaxies selected according to either Stellar Mass (SM) or Star Formation Rate (SFR) at multiple abundances and at $z=0$ and $1$. 
We find our model describes the monopole, quadrupole and hexadecapole of the galaxy-power spectra down to scales of $k\approx 0.6 [h/$Mpc] within the accuracy of our simulations. This approach could pave the way to significantly increase the amount of cosmological information to be extracted from future galaxy surveys.

\end{abstract}


%%%%%%%%%%%%%%%%%%%%%%%%%%%%%%%%%%%%%%%%%%%%%%%%%%

%%%%%%%%%%%%%%%%% BODY OF PAPER %%%%%%%%%%%%%%%%%%

\section{Introduction}

Since new observational campaigns are about to map out the distribution of galaxies with an unprecedented accuracy (e.g. Euclid, \citealt{Laurejis2011}, and \citealt{Euclid}, and DESI, \citealt[][]{DESI}), the quest for accurate predictions of the distribution and clustering of galaxies has become of great importance. In order to fully exploit these new datasets and extract the cosmological parameters of our Universe, an accurate modelling of galaxy clustering statistics, with all its features, such as e.g. redshift space distortions \citep[e.g.][]{1987MNRAS.227....1K} and baryonic acoustic oscillations \citep[e.g.][]{1998ApJ...496..605E}, is required to achieve unprecedented accuracy \cite[see e.g.][]{2013PhRvD..88f3537D,2013MNRAS.432.1928T,2015MNRAS.454.4326P}.

However, making accurate predictions of the distribution of galaxies (as they will be observed in modern galaxy surveys) is quite challenging. In principle this requires simulating the formation of structures all the way from the very early, almost homogeneous universe, up to the galaxies that we observe today (involving a large variety of different physical processes). However, in a simplified approach, one can separate the relevant large scale effects (which are mostly coming from gravity) from the small scale effects that come from baryonic processes and capture most of the consequences of the complicated galaxy formation physics into a set of flexible ``bias'' parameters.

In this approach the modelling problems can be roughly categorised into three groups: Firstly, it is required to follow the non-linear evolution of cosmic structures through gravity with an accurate treatment of the underlying dark matter distributions to reproduce small structures. Secondly, ``biased'' galaxies have to be mapped to the underlying Dark Matter (DM) distribution. Thirdly, the theory has to model the fact that measurements are affected by Redshift Space Distortions (RSD) coming from the non-linear velocity component in the line of sight of the observation. Each of these points will be discussed in more detail below, and it is the goal of this article to lay the foundations for a model that is able to handle all  of these problems to an accuracy that would be sufficient for extracting unbiased cosmological information from a Euclid-like galaxy survey.

The evolution of dark matter particles under gravity from some given initial conditions is a well posed problem which can be solved with high accuracy and efficiency with modern $N$-body codes (see e.g. \citealt{Springel2005}, \citealt{HABIB201649}, \citealt{KuhlenVogelsbergerAngulo2012} and \citealt{Garrison_2018}). However, these $N$-body simulations are computationally expensive and require considerable amount of CPU time and/or memory. Nevertheless, several approaches have been suggested in the literature to speed up their production, 
\citep[e.g.][]{2002MNRAS.331..587M, 2013JCAP...06..036T, 2016MNRAS.459.2327I} or to interpolate among the outputs of simulations \citep{HeitmannEtal2014,LiuEtal2018,NishimichiEtal2019,DeRoseEtal2019,GiblinEtal2019,EuclidEmu2019,WibkingEtal2019,WintherEtal2019,AnguloEtal2020,EuclidEmu2020}. 

Several approaches to the mapping between dark matter and observed tracers of the large scale structure have been studied. On one side, a full approach of modelling the behaviour of the baryonic physics composing galaxies has been attempted (see e.g. \citealt{Vogelsberger2013}, \citealt{Schaye2015}, \citealt{McCarthy2017}, \citealt{TNG}). 
However, modelling the behaviour of the baryonic component, including hydrodynamics, star and black hole formation together with feedback, in an efficient way remains a challenge. On the other side, a more agnostic approach can be taken. These  complexities can be parameterised by a series of functional dependencies, the bias expansion, informed by the symmetries of the underlying laws rather than the details of the specific processes \citep{2018PhR...733....1D}. For non-relativistic tracers these symmetries are the equivalence principle and translational, rotational and Galilean invariance. This approach has been extensively used under the assumption of perturbed dynamics (\citealt{VlahCastorinaWhite2016}, \citealt{Ivanov_2020}, \citealt{d_Amico_2020}, \citealt{Colas_2020}). Recently, we explored its consequences under $N$-body computed dynamics  in \cite{ZennaroAnguloPellejero2021} - see also \citealt{Kokron2021} - where we built on the ideas by \cite{Modi_2020} and built an emulator modelling biased tracers in real space over the cosmological space. This methodology finds sub-percentage accuracy in the real space power and cross spectra until scales of $k\approx 1 [h/$Mpc]. This model was recently used on observations by \cite{hefty}, but with no velocity modelling.

A vital step to obtain survey-like observations requires the treatment of peculiar velocities. Modern surveys typically observe the Universe in the three dimensional redshift space spanned up by two angular coordinates and one redshift coordinate. The latter encodes a mixture of positional information (due to the distance dependence of the gravitational redshift of observed objects) and velocity information through ``redshift space distortions'' which are caused by the Doppler effect from the peculiar velocities of tracers. From a theoretical ground, if the position and a velocity field is known for a given set of tracers, the distribution in redshift space can be predicted easily. However, modelling the velocity of biased tracers faces similar issues as the modelling of the positions: the non linear evolution due to gravity and a non linear mapping between the dark matter fluid and the velocities of bias tracers.  Since the velocity field departs from its linear theory prediction on extremely large scales ($k> 0.03[h/$Mpc]), models beyond linear theory must be used to extract cosmological information from redshift surveys. This has been long recognised and a variety of methods have been attempted to model the distortions (e.g. \citealt{Fisher1995}, \citealt{Hamana2003}, \citealt{Scoccimaro2004}  \citealt{Seljak_2011}, \citealt{JenningsBaughPascoli2011}, \citealt{BianchiPercivalBel2016}, \citealt{KuruvillaPorciani2018}, \citealt{Cuesta2020}, \citealt{Chen:2020zjt}). Moreover, strong dependence on halo bias, primarily due to the non-linear mapping between real and redshift space has already been found and broadly studied (see e.g. \citealt{TaruyaAtsushiNishimichi2010}).  A standard approach to include small-scale non-linearities, manifested as fingers-of-god, is to use a streaming model (see e.g. \citealt{ReidWhite2011} and \citealt{VlahCastorinaWhite2016}) in configuration space, where linear theory is spliced together with an approximation for random motion of particles in collapsed objects. This arises from ignoring the scale-dependence of the mapping between real and redshift space separations and assuming an isotropic velocity dispersion. This leads to a convolution of the linear theory result with a line-of-sight smearing or equivalently a multiplication of the power spectrum by the Fourier transform of the small-scale velocity probability distribution function \citep{Peacock1992}. It is also worth mentioning the works on Edgeworth streaming models from \cite{UhlemannKoppHaugg2015} and the split densities approach from \cite{Paillas2021}.  A more recent approach based on Lagrangian Perturbation Theory (LPT) can be found in the \texttt{velocileptors} code by \cite{Chen:2020zjt}, where they used one-loop perturbation theory with effective corrections for small scale effects to model the velocity statistics building on previous works in configuration space combining velocity statistics and the correlation function in LPT. 

When comparing against observations, 5-10 percent level of accuracy down to the scales of $\approx$25 [Mpc$/h$] on the correlation function (roughly $k\approx [0.3h/$Mpc] on the power spectrum) has previously been achieved in using RSD to measure the growth of structure and testing theories of gravity with galaxies from large surveys such as BOSS (see e.g. \citealt{BOSS2017}, \citealt{Chuang_2017}, \citealt{Pellejero-Ibanez_2017} and \citealt{Beutler2017} ), eBOSS (see .g. \citealt{eBOSS2020}), VIPERS (\citealt{VIPERS2017}), GAMA (\citealt{GAMA2013}), the 6 degree Field Galaxy Survey 6dFGS (\citealt{6dF2012}), the Subaru FMOS galaxy redshift survey (\citealt{Subaru2016}) and the WiggleZ Dark Energy Survey (\citealt{WiggleZ2011}). The next generation of spectroscopic redshift surveys such the Dark Energy Spectroscopic Instrument (DESI, \citealt{DESI}) and Euclid (\citealt{Laurejis2011} and \citealt{Euclid}) promise percent-level accuracy for parameter constraints, demanding unprecedented precision in the modelling.

In this paper we propose and test a new model for the clustering of biased tracers in redshift space. This model corresponds to a second-order perturbative Lagrangian bias expansion which is advected to Eulerian space employing a displacement field extracted from $N$-body simulations. Eulerian coordinates are then transformed into redshift space by directly employing simulated velocity fields augmented with nuisance parameters capturing various possible satellite fractions and intra-halo small-scale velocities. We quantify the accuracy of our approach against samples of physically-motivated mock galaxies selected according to either Stellar Mass (SM) or Star Formation Rate (SFR) at multiple abundances. 

We first explore the performance of our redshift-space model by fixing the bias parameters to their best-fit values obtained from real-space statistics (galaxy power spectrum and matter-galaxy cross power spectrum). In this way, we isolate the contribution of our velocity fields. We find that multipoles of the redshift-space power spectrum can be recovered to very high precision, with the main limiting factor being the halo-mass dependence of the intra-halo velocities. In a second set of tests, we simultaneously fit all the free-parameters of the model, including those of the bias expansion, and show that the monopole, quadrupole and hexadecapole of the galaxy power spectrum can be recovered on all scales down to $k\sim0.6 [h/\rm{Mpc}]$ with a precision better than the statistical uncertainties of the volume of the simulation. These results suggest that our model should deliver unbiased cosmological constraints employing all scales robustly measured by future large-scale structure surveys.

We divide this paper as follows. In \S\ref{sec:catalogues} we provide details of the set of mock galaxy catalogues with which we will validate our model and introduce the $N$-body simulation that will be used for the theoretical model. In \S\ref{sec:model} we discuss our model for the redshift-space clustering of biased tracers. In \S\ref{sec:fitting} we show the fitting procedure. In \S\ref{sec:results} we first validate our redshift-space model by fixing the bias parameters to those obtained in real space. We demonstrate the high accuracy of our approach and identify the mass-dependent velocity dispersion as the main limiting factor. Subsequently, we show that the multipoles of the galaxy power spectrum can be simultaneously reproduced, which indicates that our approach should return unbiased cosmological constraints. We conclude and summarise our results in \S\ref{sec:conclusions}.

\section{Galaxy-like samples}
\label{sec:catalogues}

In this section we present the simulation used throughout the work and describe how we obtain the set of galaxy mocks, to later test whether our bias model is able to fit the clustering statistics of realistic galaxies.

\subsection{Simulation}
\label{sec:simulation}
The $N$-body simulations used in this work are taken from the BACCO suite, presented for the first time in \cite{AnguloEtal2020}. This is a set of $N$-body simulations run with the \texttt{L-Gadget3} code (\citealt{Angulo2012}), a lean version of \texttt{Gadget} \citep{Springel2005}, which includes a version of \texttt{SUBFIND} \citep{Springel2001} optimised, among other things, for efficient on-the-fly calculations of many halo and subhalo properties, including properties that are nonlocal either in space or in time.

The BACCO simulations are available in four different cosmologies. For each cosmology, two simulations with same initial density amplitudes but opposite phases have been run. This makes it possible to use the ``Fixed \& Paired'' technique, presented in \cite{AnguloPontzen2016}, which allows for a significant reduction of the effect of sample variance on the matter power spectrum. In particular, we expect cosmic variance to be suppressed by two orders of magnitude on scales $k < 0.1 [h \mathrm{Mpc}^{-1}]$.

All the simulations in the BACCO suite share a comoving volume $V_{\rm sim}=1.44^3 \approx 3 \, h^{-3} \, \mathrm{Gpc}^3$, in which $4320^3$ cold matter particles are evolved from the initial redshift $z_{\rm ini} = 49$ to the late-time universe. The initial positions and velocities are set by applying a second-order Lagrangian perturbation theory displacement field to a regular grid. The numerical settings of \texttt{L-Gadget3}, detailed in \cite{AnguloEtal2020}, are chosen to guarantee a 2\% convergence of the matter power spectrum at $k \sim 10 [h \mathrm{Mpc}^{-1}]$.

In this work, we use one of the cosmologies of the BACCO suite, specifically \texttt{nenya}, employing both its phase realisations to minimize large scale noise. This cosmology is characterised by a cold dark matter density parameter of $\Omega_{\rm cdm} = 0.265$ and a baryon density parameter $\Omega_{\rm b} = 0.05$, resulting in a total matter density of $\Omega_{\rm m} = 0.315$. The linear amplitude of matter fluctuations at redshift $z=0$ is $\sigma_8=0.9$, the Hubble parameter is $h=H_0/100=0.60$, and the scalar spectral index $n_{\rm s}=1.01$. Finally, \texttt{nenya} assumes as the optical depth at recombination the value $\tau = 0.0952$, no massive neutrinos, and a constant Dark Energy equation of state with $w=-1$.

Among the simulation outputs, we use a subsampled dark matter catalogue (retaining 1 every $4^3$ particles, selected according to their Lagrangian position at the simulation starting redshift, named sDM particles). Moreover, we use Friends-of-Friends dark matter haloes, defined with a linking length of 20\% of the mean intraparticle separation, whose masses are defined as the mass enclosed in spheres with overdesities exceeding 200 times the critical overdensity. Finally, we use subhaloes individuated by \texttt{SUBFIND}, for which we have stored the peak circular velocity $v_{\rm peak}$ and peak mass $m_{\rm peak}$ and other relevant properties needed for the \texttt{SHAMe} procedure described in the following part of this section. Orphan subhaloes (subhaloes fallen below the resolution limit due to mass stripping in their parent haloes) are followed by keeping track of their most bound particle.

\subsection{Stellar Mass selected galaxies}

The galaxy mocks we use here are based on the works by \cite{ContrerasAnguloZennaro2020AB, ContrerasAnguloZennaro2020}. They rely on an extended SHAM approach that proved to recover the clustering statistics of galaxies in the TNG300 magneto-hydrodynamic simulation (\citealt{10.1093/mnras/stx3040}, \citealt{10.1093/mnras/stx3304}, \citealt{10.1093/mnras/sty2206}, \citealt{10.1093/mnras/stx3112}, \citealt{10.1093/mnras/sty618} ) in real and redshift space. Two main samples are used in this work: Stellar Mass selected galaxies and Star Formation Rate selected galaxies. Using this method, we populate the subhaloes (and orphans) of our simulation with galaxies -- each including estimates of their SM and SFR. 

For SM, the mapping of subhaloes to galaxies is defined through an abundance matching technique which is parameterized through a set of five parameters $\{\sigma_{M_*},t_{\rm merger}, f_{\rm s}, A_{\rm c}, A_{\rm s}\}$. These parameters control for scatter in the considered abundance matching relation, a variety of processes associated with the disruption of subhaloes and galaxies and further two parameters $A_{\rm{c}}$ and $A_{\rm{s}}$ which control the level of galaxy assembly bias of central and satellite galaxies respectively. However, we refer the reader to \cite{ContrerasAnguloZennaro2020} for a detailed explanation of these parameters and for tests against realistic hydrodynamical setups and \cite{ContrerasAnguloZennaro2020AB} for the assembly bias implementation of the model. For the SM selected sample we choose the parameters  $\{ 0.1786, 0.6617, 0.0073, 0.2404, -0.1163\}$ which have been tuned to fit the TNG300 magneto-hydrodynamic simulation at the different number density cuts. To create a galaxy sample we then select the $N$ galaxies with the highest stellar mass content, where $N$ is chosen to produce a given number-density $n$.

The sample is computed at two redshifts, $z=0$ and $z=1$. It is further divided in three cuts on number densities providing $\bar{n}=0.00316$[$h/$Mpc]$^{3}$, $\bar{n}=0.001$[$h/$Mpc]$^{3}$ and $\bar{n}=0.0003$[$h/$Mpc]$^{3}$. We expect this sample to behave as what is assumed from selection criteria followed by Luminous Red Galaxies (LRG's) surveys.

\subsection{Star Formation Rate selected galaxies}

Following \cite{ContrerasAnguloZennaro2020} we assign a Star Formation Rate to each subhalo using a semi-empirical model, based on the EMERGE technique \citep{Moster2018} but optimised to be efficiently run on Dark Matter simulations in a matter of seconds, without the need of the full merger history of haloes. In a nutshell, the model assumes that the SFR is equal to the recent accreted mass of its subhalo multiplied by an efficiency function. This model has 5 free parameters to control the star-formation efficiency ($\beta$ and $\gamma$), the host mass at which the formation of stars peaks ($M_1$), and the dynamical disruption of galaxies in haloes ($\tau_0$ and $\tau_{S}$). We set these parameters to $\beta=2.49$, $\gamma=5.13$, $\log M_1=12.77$, $\tau_0=4.93$ and $\tau_S=-0.36$, which correspond to the best fitting parameters presented in Tab. 1 of \cite{ContrerasAnguloZennaro2020} when fitting the TNG300 galaxy clustering.
The sample is also computed at two redshifts, $z=0$ and $z=1$. It is further divided in three cuts on number densities providing $\bar{n}=0.00316$[$h/$Mpc]$^{3}$, $\bar{n}=0.001$[$h/$Mpc]$^{3}$ and $\bar{n}=0.0003$[$h/$Mpc]$^{3}$.

\section{The Model}\label{sec:model}

Our model is based on a $N$-body simulation, which is used to map a biased galaxy density field from Lagrangian space into redshift space through a set of bias parameters. The main relevant components that are used from the $N$-body simulation are the displacement of the dark matter particles and the velocity field of the subhaloes. These exactly encode all of the gravitational evolution for dark matter particles, but also approximately encode most of the spatial gravitational evolution of galaxies. Galaxies and dark matter are expected to have almost the same displacement field on large scales \citep{2018PhR...733....1D}. On small scales these two diverge due to baryonic processes on the one hand, and due to the fact that galaxies come from an ensemble of dark matter particles on the other hand \citep{Arico2020}. This makes it possible to handle the questions of where galaxies form (depending on the initial environment) and at what spatial locations galaxies end up (depending on the displacement field) separately.

Therefore our model operates by: 1) constructing a galaxy density field in Lagrangian space as a function of the initial linear fields of the simulation through a small set of flexible bias parameters, 2) advecting this field to Eulerian space by using the non-linear displacement field from the $N$-body simulation, and 3) imposing RSD by using a combination of the velocity field of the $N$-body simulation and two additional RSD parameters.

We will use for the model the same $N$-body simulation presented in \S\ref{sec:simulation}. This allows to further reduce the impact of cosmic variance on the measurements because both model and galaxy mocks will share the same initial seeds.

\subsection{Bias mapping} \label{sec:biasmapping}
\begin{figure*}
    \centering
    \includegraphics[width=\textwidth]{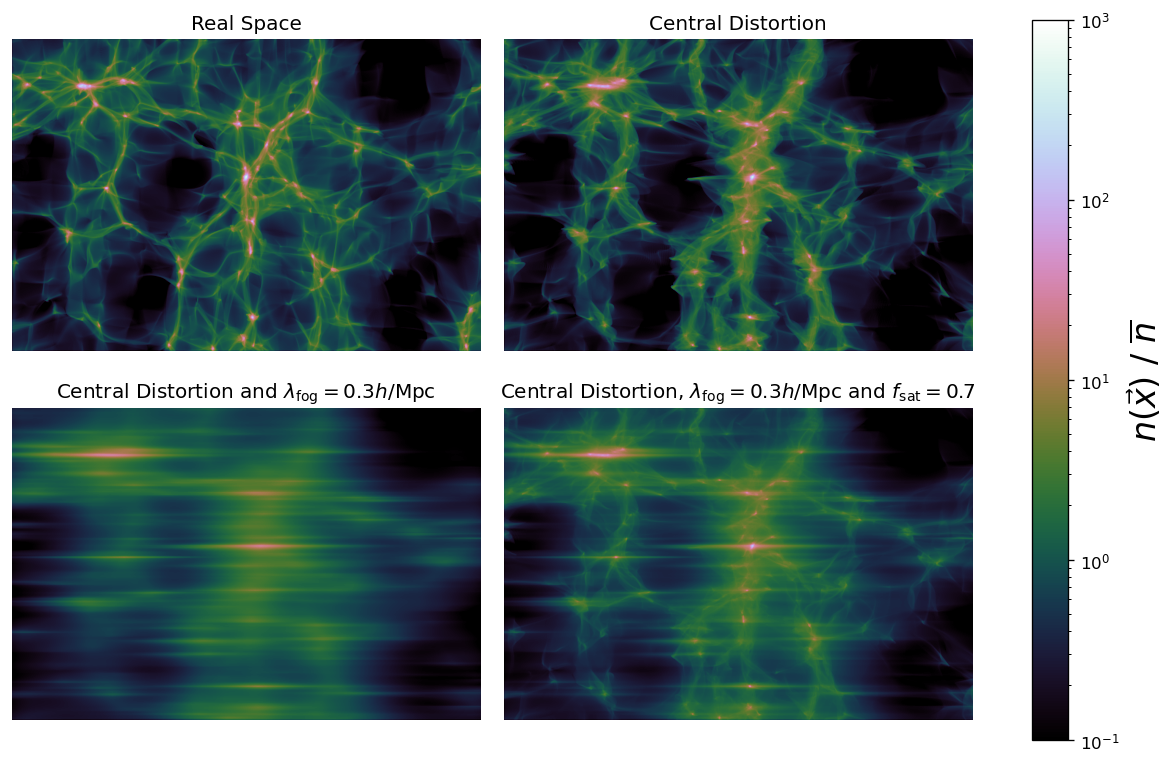}
    \caption{Projected density fields that illustrate the steps that are used for obtaining the clustering in redshift space. Top Left: A biased field in real space (created as described in section \ref{sec:biasmapping}). Top Right: The same field, but with the combined centre of mass RSD applied as described in section \ref{sec:redshiftspacedistortions}. Bottom left: The displacement distortion field, but with a FoG applied as a 1D Exponential smoothing. Bottom right: a combination of the two latter fields with a satellite fraction $f_{\rm{sat}} = 0.7$ as in Eq.~\ref{eq:FoG}, but in real space. The projected volume has a width of 64[Mpc$/h$] and a depth of 20[Mpc$/h$].}
    \label{fig:fog_fields}
\end{figure*}

We will work within the context of second order Lagrangian bias (\citealt{PhysRevD.78.083519}). Thus, first we apply our bias model onto the linear density fields of the simulation defining a bias field on the Lagrangian space coordinates $\pmb{q}$ and then move those tracers to Eulerian space coordinates $\pmb{x}$ using the displacement field of the simulation similarly to what was done in \cite{Modi_2020} and \cite{ZennaroAnguloPellejero2021}. The symmetries we use up to second order in bias expansion define the following model for the overdensity of tracers $\delta_{\rm{tr}}(\pmb{q})$ in Lagrangian space,
\begin{equation}
\begin{split}
    \delta_{\rm{tr}}(\pmb{q}) = & 1 + b_1\delta_{\rm{L}}(\pmb{q}) + b_2 \left( \delta^2_{\rm{L}}(\pmb{q})-\langle\delta^2_{\rm{L}}(\pmb{q})\rangle \right) \\ & + b_s \left( s^2(\pmb{q}) - \langle s^2(\pmb{q})\rangle \right) + b_{\nabla} \nabla^2 \delta_{\rm{L}}(\pmb{q}) \; .
	\label{eq:model}
\end{split}
\end{equation}

Here $\delta_{\rm{L}}(\pmb{q})$ stands for the linear field as given by the simulation and $s^2$ is the traceless part of the tidal field $s^2=s_{ij}s^{ij}=(\partial_i\partial_j\phi(\pmb{q}) - 1/3\delta^{\rm{K}}_{ij}\delta_{\rm{L}}(\pmb{q}))^2$ with $\phi(\pmb{q})$ the linear displacement potential in Lagrangian coordinates, $\nabla^2\phi = 4\pi G \bar{\rho}\delta_{\rm{L}}(\pmb{q})$.

For creating the tracer field we follow the procedure described in \cite{PhysRevD.100.043514} and \cite{ZennaroAnguloPellejero2021}. We use the uniform grid of pseudo particles in Lagrangian space which we refer to as tracer particles. These can in principle be chosen to be the same set of particles as the ones of dark matter particles in the simulation, but we found that it is often sufficient to use only a subset of the particles (e.g. 64 times less) to convergedly resolve structures up to the scales we are interested in. For each of the tracer particles, the biased Lagrangian density field $\delta_{\rm{tr}}$ (or its individual components) are evaluated at their Lagrangian coordinates. The Eulerian positions $x$ are taken from the simulated particle which is closest in Lagrangian space (which will coincide with the sDM particles, see \S\ref{sec:simulation}).

Then we deposit the particles to a grid in Eulerian space with a weight that is given by  Eq.~(\ref{eq:model}) through a CIC assignment. In principle, when given a set of bias parameters, one could directly weight by $\delta_{\rm{tr}}$ as defined in  Eq.~\eqref{eq:model}.  However, since we want to fit and test a large range of bias parameters, we instead create five separate Eulerian fields that are weighted each by only one of the components in Eq.~\eqref{eq:model} and that can later be linearly combined to create the galaxy field for any set of bias parameters. For example for the case of the ``1'' term the weight is 1 for all particles and the resulting field corresponds to the dark matter density field. For more details see \citet{ZennaroAnguloPellejero2021}. 

From the resulting biased tracer fields in Eulerian coordinates, we extract the clustering statistics to compare with the galaxy samples.

\subsection{Redshift Space Distortions} \label{sec:redshiftspacedistortions}

In this section, we extend our implementation into RSD. Since we have access to the full non-linear $N$-body simulations, we are not limited to only use perturbation theory solutions as previous recent approaches have (eg. \citealt{2020arXiv201203334S}).

In redshift space the clustering is not isotropic, but different in the line of sight direction (chosen here as the $z$-direction) than in the other two directions. Therefore, the power spectrum is defined in redshift space through two parameters: $k$ the magnitude of the wave-vector, and $\mu = k_z / k$ the relative amplitude of the wavevector in the redshift direction. The 2D redshift space power spectrum is then defined through
\begin{align}
    \langle \delta(k,\mu)\delta(k^{\prime},\mu^{\prime}) \rangle = 2\pi \delta_{D}(k-k^{\prime})\delta_{D}(\mu-\mu^{\prime})P(k,\mu) \; .
\label{eq:2D_pk}
\end{align}

To obtain the redshift space density field $\delta(\pmb{s})$, we deposit particles on a 3D grid where the position of the tracer particles are shifted over the line of sight ($\hat{\pmb{z}}$) by 
\begin{equation}
    \pmb{s}_{\rm{tr}}=\pmb{x}_{\rm{tr}}+\frac{\hat{\pmb{z}} \cdot \pmb{v}_{\rm{tr}}(\pmb{x})}{aH} \, \hat{\pmb{z}} \; ,
	\label{eq:RSD}
\end{equation}
where we need to define $\pmb{v}_{\rm{tr}}(\pmb{x})$ based on the velocities provided by the simulation. The most obvious solution would be to use the velocity of the nearest dark matter particle in Lagrangian space (analogously to our choice for the positions of the tracer). However, this generally overestimates the velocity dispersion of galaxies. For example, central galaxies will have a much smaller velocity dispersion than the typical velocities of dark matter particles that constitute the halo the galaxy belongs to. This is so since such a galaxy is formed from a much larger Lagrangian volume than a single dark matter particle. This larger Lagrangian volume implies a larger amount of particles whose velocities will be averaged. The more velocities are averaged, the less the velocity dispersion with respect to dark matter particles one. Therefore it is, a priori, not possible to infer the correct velocity distribution of a selected set of galaxies, just given the dark matter distribution.

Therefore we attempt here to define a model which can flexibly adapt to the velocity statistics of different sets of galaxies. We split the redshift space distortions into two components. The first component we refer to as \emph{central distortion} which models the redshift space distortions due to the velocities of whole clusters of galaxies, but explicitly excludes the effect of the intra-cluster velocities of galaxies. On top of this central distortion-field, we then model intra-cluster velocities (which cause the famous Finger of God effect) through effective smoothing parameters that can adaptively be fitted to any set of galaxies.

\subsubsection{Central distortion}
The ``central-distortion'' effect can be easily and uniquely inferred from a given $N$-body simulation, and does not depend on the selected population of galaxies: each particle that is part of a halo gets mapped into redshift space not with its own velocity, but instead with the velocity of the central subhalo of the same halo. Particles lying outside of haloes get simply mapped with their own velocity:

\begin{equation}
\pmb{v}_{\rm{tr}} = \begin{cases} \pmb{v}_{\rm{DM}} , & \mbox{if the tracer is outside of a halo} \\ \pmb{v}_{\rm{main\,sub}}, & \mbox{if the tracer is inside of a halo} \end{cases}
\label{eq:velocity}
\end{equation}
In this way, all particles of the same halo get mapped into redshift space with the same velocity, effectively removing the Finger of God effect from the redshift space projection.  We display an example of the obtained redshift space distortions in the top right panel of Figure \ref{fig:fog_fields}. We expect this to work well for the redshift space distortions of central galaxies. However, for satellite galaxies we explicitly removed their intra-cluster velocity dispersion, and therefore the central distortion field exhibits no FoG effect.
Note this distortion roughly accounts for the so called ``Kaiser effect'' \citep{1987MNRAS.227....1K} but there are extra effects coming from the specifics of the main subhaloes formation and also accounts for the non-linearity in the large scale velocity flows.

\subsubsection{Including velocity dispersion}

The FoG effect is traditionally included as a convolution with a Gaussian function \citep{Peacock1992} or an exponential function \citep{1983ApJ...267..465D} in real space. The first approach seems to work well for samples separated in mass bins \citep{10.1093/mnras/stx3349} but the second is more realistic since galaxies populate
haloes of a wide range of masses and velocity dispersions \citep{White2001}. This can be thought as a combination of several Gaussian distributions leading to an exponential behaviour. We test the performance of both models on the Probability Distribution Function (PDF) of relative velocities of the galaxy catalogues in Fig~\ref{fig:PDF_relvels} at $\bar{n}=0.00316[h/$Mpc$]^3$. About 26 \% of the galaxies in this catalogue reside in central subhaloes and have therefore by our definition exactly zero velocity. We excluded these from the histogram, but they can easily be pictured as a Dirac delta function peak at 0 containing $\approx 26\%$ of the total area in Figure \ref{fig:PDF_relvels}.

We find that the exponential function fits the velocity distribution generally better. That means the velocity distribution of galaxies relative to central subhaloes can be well described by
\begin{align}
  p(v_z) &= (1 - f_{\rm{sat}}) \delta_{\rm{D}} (v_z) + f_{\rm{sat}} \exp \left( -\lambda v_z \right) \; .
\end{align}
Where  $\delta_{\rm{D}}$ is the Dirac-delta distribution which accounts for the fact that a large fraction of the galaxies ($1 - f_{\rm{sat}}$) are central galaxies having, by definition, zero velocity with respect to themselves. Here $f_{\rm{sat}}$ refers to the satellite fraction and $\lambda_{\rm{FoG}}$ models the amount of damping due to peculiar velocities as discussed in \cite{10.1093/mnras/stx3349} (although in their work a Gaussian function was used). Therefore we can model the effect of the intra-cluster velocities onto the redshift-space distribution by applying a convolution on top of the central-distortion field along the $z$-direction:

\begin{figure}
    \centering
    \includegraphics[width=\columnwidth]{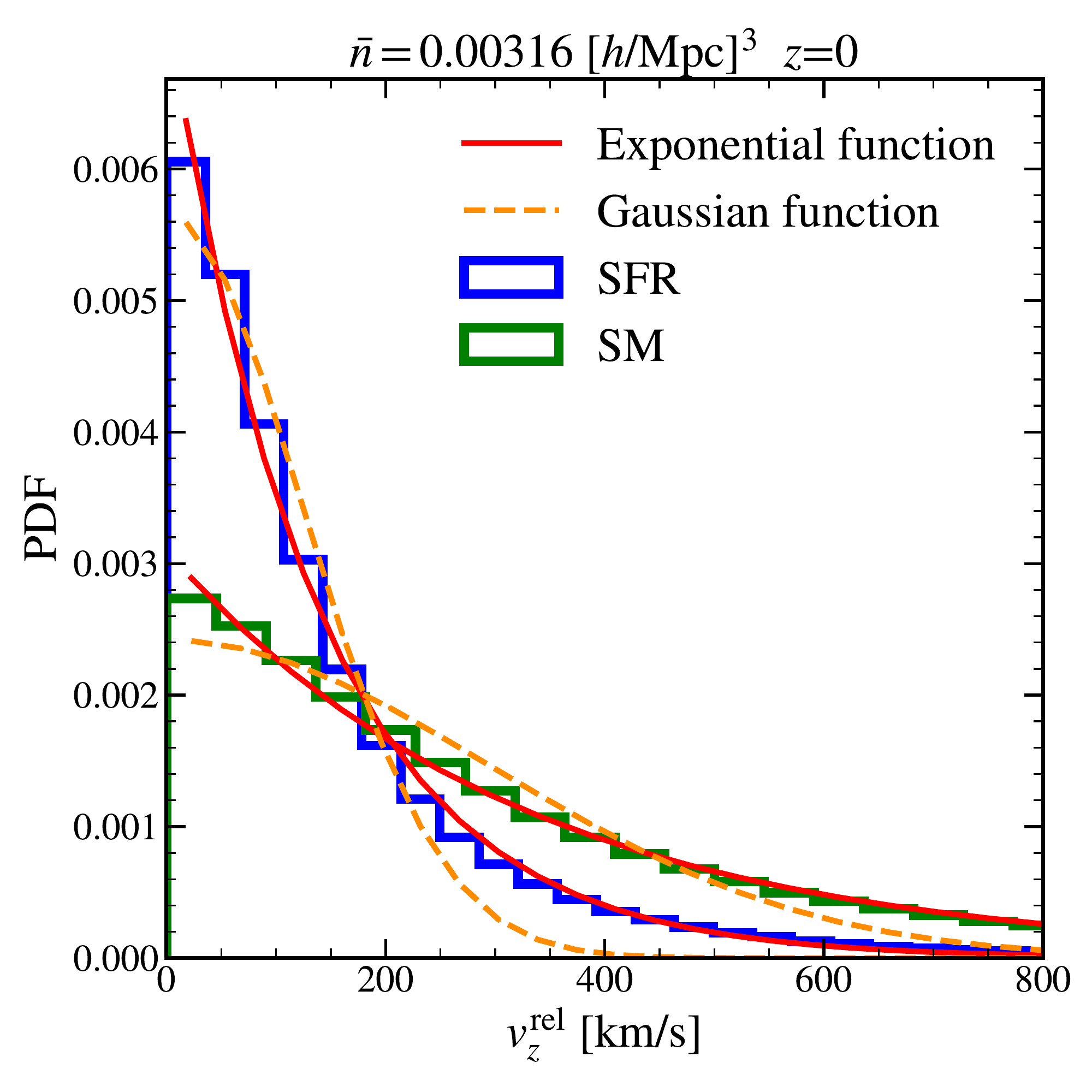}
    \caption{Probability Distribution Function (PDF) of the relative velocities for the galaxy mocks presented in \S\ref{sec:catalogues}. Both a Gaussian and an Exponential model are fitted to the distribution. The exponential model shows a better fit to the whole range of relative velocities.}
    \label{fig:PDF_relvels}
\end{figure}

\begin{equation}
\delta^{\rm{FoG}}_{\rm{tr}}= \delta_{\rm{tr}} \boldsymbol{\ast}_z \left( (1-f_{\rm{sat}}) \delta_{\rm{D}}(s_z) +f_{\rm{sat}}\exp \left( {-\lambda_{\rm{FoG}} s_z} \right) \right) \; .
\label{eq:densFoG}
\end{equation}

Note that $\lambda_{\rm{FoG}}$ is a free parameter (with units [$h/$Mpc]) which is not directly the dispersion of peculiar velocities (it is related to it by its inverse). We use the notation $\boldsymbol{\ast}_z$ for the 1D convolution operation in the $z$-direction. Going to Fourier space and computing the power spectrum (Eq.~\ref{eq:2D_pk}) we end up with,

\begin{equation}
P_{\rm{tr}}^{\rm{FoG}}(k,\mu) = P_{\rm{tr}}(k,\mu)\left( (1-f_{\rm{sat}})+ f_{\rm{sat}} \frac{\lambda^2_{\rm{FoG}}}{\lambda^2_{\rm{FoG}}+k^2\mu^2} \right)^2 \; .
\label{eq:FoG}
\end{equation}

Both $f_{\rm{sat}}$ and $\lambda_{\rm{FoG}}$ parameters can be treated as nuisance parameters when fitting the clustering in redshift space. The motivation for this expression is that central galaxies are well described by our tracer velocities, while we model the velocity dispersion of satellite galaxies through an Exponential distribution with fixed width. Which fraction of a galaxy sample are satellite or central galaxies will of course depend on the selection criterion of that galaxy sample and therefore it is useful to have the relative contributions as the free fitting parameter $f_{\rm{sat}}$. In the bottom panels of Figure \ref{fig:fog_fields} we illustrate how these operations would look in Eulerian space.

Furthermore, we will include a noise term treated as a free amplitude Poissonian noise in Eulerian space. This noise will only contribute to the monopole of the 2D power spectrum in Eq.~\ref{eq:FoG} as a constant spectrum of size $A_{\rm{noise}}\times \bar{n}^{-1}$,  where $\bar{n}$ refers to the number density of interest and $A_{\rm{noise}}$ is a nuisance parameter included to account for the amplitude of the noise. Thus, if $A_{\rm{noise}}$ is greater (lower) than one we will be talking about overpoissonian (subpoissonian) behaviour of the noise. How the noise is defined and evolves from Lagrangian to Eulerian space is still discussed and the model presented here is simply an heuristic approach.  
%Note that this same procedure could have been done at the field level with a convolution in real space with the second term of the previous equation but our goal is to fit the clustering of sample galaxies, therefore, we optimise every step. \js{mention that these are two new free parameters}

\subsection{Optimisation}

Our goal in this paper is to test how well this model based on second order bias expansion reproduces the two-point clustering of galaxy-like samples. As a consequence, the free parameters $\{ b_1, b_2, b_s, b_{\nabla}, f_{\rm{sat}}, \lambda_{\rm{FoG}}, A_{\rm{noise}} \}$ need to be found for each of the samples. For this reason we optimise the procedure so that we do not compute $\delta_{\rm{tr}}$ from the simulation per set of bias parameters. We exploit the linearity of all operations done within the context of expansion bias. Therefore, we compute the 2D power/cross-spectra for each of the 15 combinations of the terms of Eq.~\ref{eq:model} and then construct the $P_{\rm{tr}}(k,\mu)$ by linearly combining these, weighted appropriately by a given set of bias parameters. Finally, we construct $P_{\rm{tr}}^{\rm{FoG}}(k,\mu)$ from Eq.~\ref{eq:FoG}. This will allow us to make the fits using usual minimisation techniques.

\section{Fitting procedure}
\label{sec:fitting}

When fitting the real space clustering, we use a Gaussian likelihood function as the difference between galaxy-mocks and model, weighted by a Gaussian field covariance matrix (following \citealt{ZennaroAnguloPellejero2021}). In this section, we extend the previous approach to fit data in redshift space.

From our galaxy mocks and from our bias-models we can easily compute these power spectra and could in principle directly fit the bias-model spectra to the mock 2D-redshift space spectra as defined in Eq.~\ref{eq:2D_pk}. However, it is simpler to decompose the 2D-redshift space power spectrum into a set of 1D-functions through a multipole expansion. From a given 2D-redshift space spectrum we calculate the multipole spectra as
\begin{equation}
P^\ell(k)=\frac{2\ell+1}{2}\int_{-1}^{1}d\mu\,P(k,\mu)\mathcal{P}_\ell(\mu),
\label{eq:pkpoles}
\end{equation}
where $\mathcal{P}_\ell(\mu)$ is the Legendre polynomial of order $l$.

We fit these Legendre multipoles of the galaxy power spectrum via the maximisation of a likelihood given by
\begin{multline}
 \log \LL = -\frac{1}{2} \left(P^\ell_{\rm gal}(k) - {P^\ell_{\rm model}}(k)\right)^{\rm{T}}\,C_g^{-1}\,\left(P^\ell_{\rm gal}(k) - {P^\ell_{\rm model}}(k)\right) \\ - \frac{1}{2}\log|C_{\rm{g}}|- \frac{N}{2}\log(2\pi) \, , 
\label{equation:clust_like}
\end{multline}

\noindent where $P^\ell_{\rm gal/model}$ can be given through each of the Legendre multipoles from above. We fit up to the hexadecapole, $\ell=4$. Since the dipole $\ell=1$ and tripole $\ell=3$ moments  have to be zero due to symmetry considerations this leaves us with three 1D functions which we will plot throughout the following section.

We describe the covariance matrix of the mock data, $C^g[P(k),P(k')]$, in linear theory assuming Gaussianity. Note that it is not correct from the theoretical perspective since it does not include any of the non-linearities associated to our problem. Nevertheless, a complete description of such a covariance matrix all the way up to our scales of interest ($k\approx 0.7 [h/$Mpc]) is beyond the scope of this work. Hence, we use this approximation for the fits. Explicitly, we assume that $P(k,\mu)$ follows a Gaussian distribution which leads to the following relation mode-by-mode covariance (see \citealt{Feldman:1993ky} for the monopole case),

\begin{equation}
 \label{eq:ps_cov}
 {\rm{C^g}}\left[P(k), P(k^\prime)\right] = \frac{2 (2 \pi)^3}{V} \delta_\mathrm{D}(k - k^\prime) \left[ P(k, \mu) + \bar n^{-1} \right]^2,
\end{equation}

\noindent where $V$ and $\delta_\mathrm{D}$ stand for the volume of the sample and the Dirac delta respectively. Consequently, the multipole covariance matrix is \citep{Grieb:2015bia} :

\begin{equation}
 \label{eq:ps_ell_cov_bin}
 C^g_{{\ell_1}{\ell_2}}(k_i,k_j) = \frac{2 (2 \pi)^4}{V_{k_i}^2} \delta_{ij} \int_{k_i-\Delta k/2}^{k_i+\Delta k/2} \sigma^2_{\ell_1\ell_2}(k) k^2 {\rm{d}} k \, ,
\end{equation}

\noindent where the volume of the shell in $k$-space is $V_{k_i} = 4 \pi [(k_i + \Delta k/2)^3 - (k_i - \Delta k/2)^3] / 3$ and 

\begin{multline}
 \label{eq:ps_cov_ell_ell_in_ps_ell}
 \sigma^2_{\ell_1\ell_2}(k) = \frac{2 (2 \ell_1 + 1) (2 \ell_2 + 1)}{V} \\ \times \sum_{\ell_3=0}^\infty \sum_{\ell_4=0}^{\ell_3} \left[ P_{\ell_4}(k) + \frac{1}{\bar n} \delta_{\ell_4 0} \right] \left[ P_{\ell_3 - \ell_4}(k) + \frac{1}{\bar n} \delta_{(\ell_4 - \ell_3) 0} \right] \\
 \times \sum_{\ell=\max(|\ell_1-\ell_2|,|2\ell_4-\ell_3|)}^{\min(\ell_1+\ell_2,\ell_3)} \begin{pmatrix} \ell_1 & \ell_2 & \ell \\ 0 & 0 & 0 \end{pmatrix}^2 \begin{pmatrix} \ell_4 & \ell_3 - \ell_4 & \ell \\ 0 & 0 & 0 \end{pmatrix}^2,
\end{multline}

\noindent where terms in round parenthesis represent Wigner 3j-symbols. The previous expression Eq.~\ref{eq:ps_ell_cov_bin} is assumed as the covariance for the galaxies but tells nothing about the error of the model. Even though it is generally assumed that theoretical models have no associated errors, since they are based on analytical computations that do not suffer from common sources of noise such as cosmic variance or shot noise, the truth is that theoretical predictions for a given observable are not exact but have intrinsic uncertainties. Specifically, for the case of models built from $N$-body simulations, uncertainties in the initial conditions (cosmic variance),  arbitrariness in group finders, errors introduced by the finite accuracy of  force calculation and time integration, etc, will result in stochastic predictions. Similarly, in the case of perturbation theory, uncertainty can arise from the contribution of neglected orders, approximations in the equations of motions, and also from neglected physics such as galaxy formation (see eg. \citealt{Baldauf2016}). Our hybrid model will suffer from both sources of uncertainties. It can be shown that including these ``theory'' errors is as simple as adding the theory covariance to the data covariance (see eg. \citealt{Audren2013}, \citealt{Sprenger2019} and \citealt{Pellejero-Ibanez2020}).

We then either minimise Eq.~\ref{equation:clust_like} or compute contours through the \texttt{MULTINEST}\footnote{https://github.com/farhanferoz/MultiNest} Bayesian inference tool for recovering confidence intervals ( \citealt{multinest1}, \citealt{multinest2}, \citealt{multinest3}). For the volume of the sample $V$ we use $V_{\rm sim}=1.44^3 \approx 3 \, h^{-3} \, \mathrm{Gpc}^3$, which is the volume of our simulation. Naively, since we used paired-simulations to compute the galaxy mocks, we would have used twice this volume. Nevertheless, the model itself will suffer from cosmic variance too in both density field and velocity assignment. Thus we leave the volume of the simulation as it is. The number density corresponds to the three number densities described in the previous section.

In order to give an intuition in the range $[0.5h/{\rm{Mpc}}<k<0.6h/{\rm{Mpc}}]$, these errorbars account for about 0.3\% of the monopole, 3\% of the quadrupole and 6\% of the hexadecapole's signal at a number density of 0.00316[$h/$Mpc]$^{3}$; 0.3\% of the monopole, 5\% of the quadrupole and 10\% of the hexadecapole's signal at a number density of 0.001[$h/$Mpc]$^{3}$; 0.4\% of the monopole, 10\% of the quadrupole and 20\% of the hexadecapole's signal at a number density of 0.0003[$h/$Mpc]$^{3}$. We do not phrase the results accuracy in these terms because when the signal approaches zero, these numbers lose interpretability.

\section{Results}
\label{sec:results}

In this section we test the consistency and accuracy of the model proposed in this article when applied to the galaxy samples presented in \S\ref{sec:catalogues}. We do this in two steps.

In \S\ref{sec:realtoredshift} we keep the value of all bias parameters fixed to the values that can be found through a pure real-space fitting procedure as we presented in \cite{ZennaroAnguloPellejero2021}. We then check how well the redshift power spectra are reproduced when subsequently adding the different components of our redshift space model as represented in Fig.~\ref{fig:fog_fields}. In this way, we show that the proposed parameters for modelling the redshift space distortions indeed carry the meaning that we expect and that there exist no major degeneracies between the bias parameters and the redshift space parameters.

\S\ref{sec:fitredshift} shows the main results of this work, the fitting of redshift space clustering.

\subsection{From real to redshift space} \label{sec:realtoredshift}

The model has two distinct parts; one involved in real space predictions, with $\{ b_1, b_2, b_s, b_{\nabla} \}$ as relevant parameters, and another affecting redshift space predictions, with $\{ \lambda_{\rm{FoG}}, f_{\rm{sat}} \}$. We consider $A_{\rm{noise}}$ to be a nuisance parameter throughout the rest of the work and let it free for every case\footnote{For a better understanding of how the noise enters into RSD analysis we refer to \cite{2020arXiv201203334S}. There the authors showed that the analysis includes $\mu$ dependent noise terms. We do not take them into account here.}. In this section we ask whether the two redshift-related parameters capture the correct dependencies in redshift space, and if they are both needed to accurately fit the bias model to the clustering statistics of surveys that contain a variety of different selection criteria.

To test this, we first fit the bias parameters $\{ b_1, b_2, b_s, b_{\nabla} \}$ to the real space clustering. With those parameters fixed, we then reconstruct the redshift space multipoles, but using different components of the redshift space model: 1) using only the ``central distortion'' component, 2) using the central distortion plus a fit for $\lambda_{\rm{FoG}}$ (with $f_{\rm sat} = 1$ fixed) and 3) using the full redshift space model with a fit for $\lambda_{\rm{FoG}}$ and $f_{\rm sat}$. In this way we can distinguish what is the information carried by each of the parameters. 

\begin{figure*}
    \includegraphics[width=\textwidth]{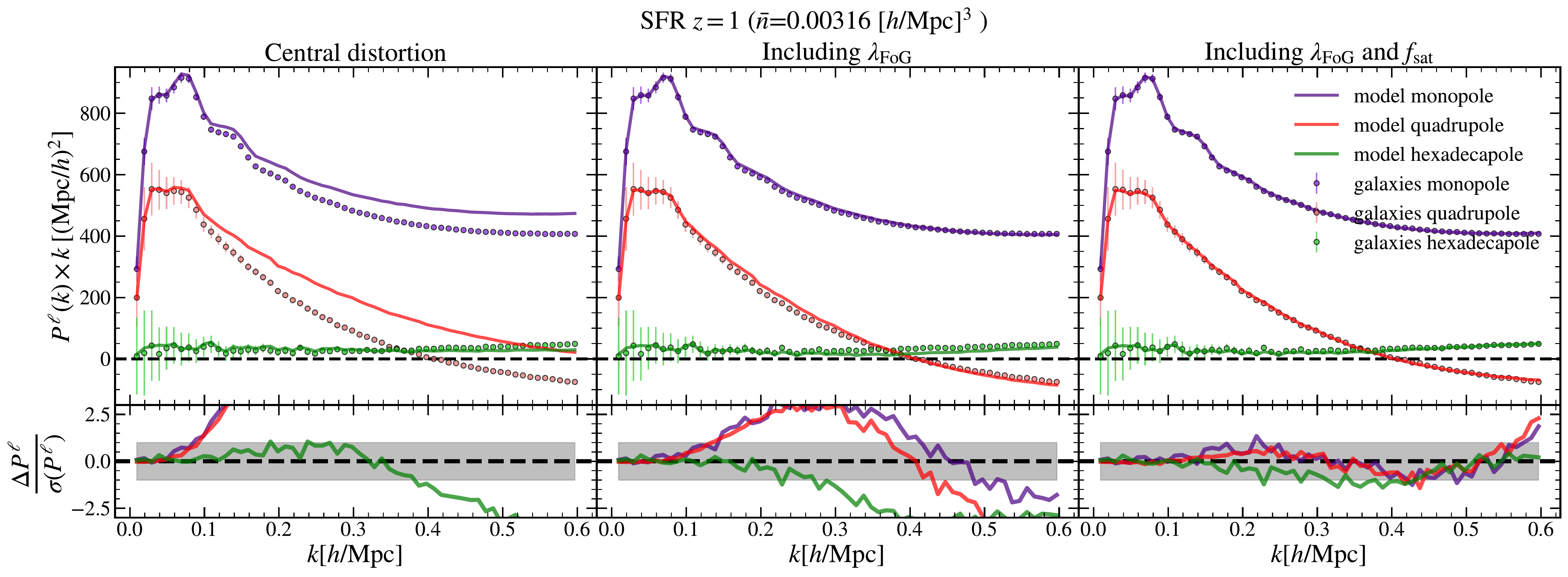}
    \caption{Reconstructed redshift space power spectra when turning on different components of our redshift space model, while keeping the real space model fixed. The different lines correspond to the modelled monopole, quadrupole and hexadecapole of the power spectrum in redshift space for SFR selected galaxies. Dots represent the same statistichs but for the target galaxy samples. Here we use the bias parameters measured from fitting power spectrum and cross-spectrum in real space $\{ b_1, b_2, b_s, b_{\nabla} \}$. Upper panels show the overall fit. Lower panels show the difference between model and data power spectra weighted by the square root diagonal terms of Eq.~\ref{eq:ps_cov}. The grey region represents the 1-$\sigma$ region assuming Gaussian noise and a volume of ~3$h^{-3}$Gpc$^3$. Note also that $A_{\rm{noise}}$ is a nuisance parameter in every case.}
    \label{fig:Validation}
\end{figure*}

The results of directly using real space fitted parameters on the SFR selected sample, at $z=1$, to a number density of $\bar{n}=0.00316[h/$Mpc$]^3$ are shown in Fig.~\ref{fig:Validation}. This Figure shows, from left to right, the increase need to account for both, the Fingers of God effect $\lambda_{\rm FoG}$ and the satellite fraction $f_{\rm{sat}}$ on the densest number density studied here. Accounting for the central distortion velocities is not enough for reproducing clustering in redshift space (left most panel). The middle panel shows how much the fit improves if a free Exponential finger of God effect,  $\lambda_{\rm{FoG}}$, is included but still exhibits clear departures from the target clustering. The right most panel shows the effect of including both $\lambda_{\rm{FoG}}$, and the fraction of satellites, $f_{\rm{sat}}$. With this incorporation, the target clustering is recovered to lie within the errorbars. 

We chose this number density since it was the most stringent and the least affected by shot-noise, but we checked all three number densities at both $z=0$ and $z=1$ and found that the level of improvement was similar or greater than at $z=1$. As a side note, in the cases at $z=0$, the use of $f_{\rm{sat}}$ does not provide an improvement and the redshift space clustering is well described by just Eq.~\ref{eq:velocity} and a constant FoG effect, $\lambda_{\rm{FoG}}$. These results show how we can translate the information found in real space to redshift space without loss of interpretability (at this level of accuracy) just including the two extra parameters $\lambda_{\rm{FoG}}$ and $f_{\rm{sat}}$.

We would like to emphasise this result, since it is expected that Euclid will observe Emission Line Galaxies at around $z=1$ (which has a similar clustering as Star Formation Rate selected samples as shown by \citealt{Gonzalez-Perez2020} and \citealt{Jimenez2020}), meaning that our model has the potential to describe a Euclid-like sample two-point statistics all the way down to scales of $k\approx 0.6 [h/\rm{Mpc}]$.

The results of directly using real space fitted parameters on the SM selected sample are left for Appendix~\ref{App:SM_shuffle}. In general, they agree well within the errorbars, but we decided to show in the left most panel of Fig. \ref{fig:Validation_SM_z0_shuffletests} the performance of the model at $z=0$, with the $\bar{n}=0.00316[h/$Mpc$]^3$ cut, because it turns as the most interesting case. It shows a clear inconsistency of the model in the mapping from real to redshift space. As discussed in the Appendix, this inconsistency is tracked to the lack of dependence of the FoG effect in Eq.~\ref{eq:RSD} with the central halo properties (such as halo mass).   

Summarising, the tests show a good level of agreement between the model and the mocks at $z=1$. Moreover, these results are kept with SM and SFR selected galaxies at $z=0$. Nevertheless, we find systematic deviations of the model for the densest SM selected galaxies at $z=0$ that we associate with the lack of modelling of the peculiar satellite velocities with its central halo properties, such as mass. This deviation gives us a good opportunity to check how the model behaves when having no information about real space bias values (which will be the case with real samples). We expect that some expansion term will try to swallow the dependencies at the two-point level and will depart from its real space counterparts to provide a better fit. We will explore this in the next section.

Note that the set of real space parameters measured in this subsection can be seen in Fig.~\ref{fig:bias_vals} as shaded regions. This Figure will be discussed in the following \S\ref{sec:fitredshift}. These real space parameters are computed following the procedure described in \cite{ZennaroAnguloPellejero2021}. In a nutshell, it consists on picking the same galaxy mock in real space and using \texttt{MULTINEST} to find the bias parameters defined in real space $\{ b_1, b_2, b_s, b_{\nabla} \}$. We would like to highlight that the Gaussian error at small scales $\sigma(0.6[h/$Mpc$])$ heavily underestimates the expected errors from non-linear predictions, since it misses all small-scales correlations.

\subsection{Redshift space} \label{sec:fitredshift}

Now we move to a full study in redshift space, where no previous information about real space is used. Here we include all parameters $\{b_1, b_2, b_s, b_{\nabla}, A_{\rm{noise}}, \lambda_{\rm{FoG}}, f_{\rm{sat}} \}$ to fit the galaxy samples clustering in redshift space. It is expected that redshift space breaks some of the degeneracies between parameters and provides an overall better fit to the galaxy multipoles. It is important then to compare with the results obtained in real space to show that the compatibility holds and thus the meaning of the model bias parameters remains. We further test this as a function of the maximum scale included in the fit, $k_{\rm{max}}$, hence also testing the consistency between measured parameters with scale. This will check if there are tensions between bias parameters fitted at different scales. Because of the relevance of these results, we split our results into SFR and SM selected galaxies. 

\subsubsection{Star Formation Rate selected galaxies results}

\begin{figure*}
    \includegraphics[width=\textwidth]{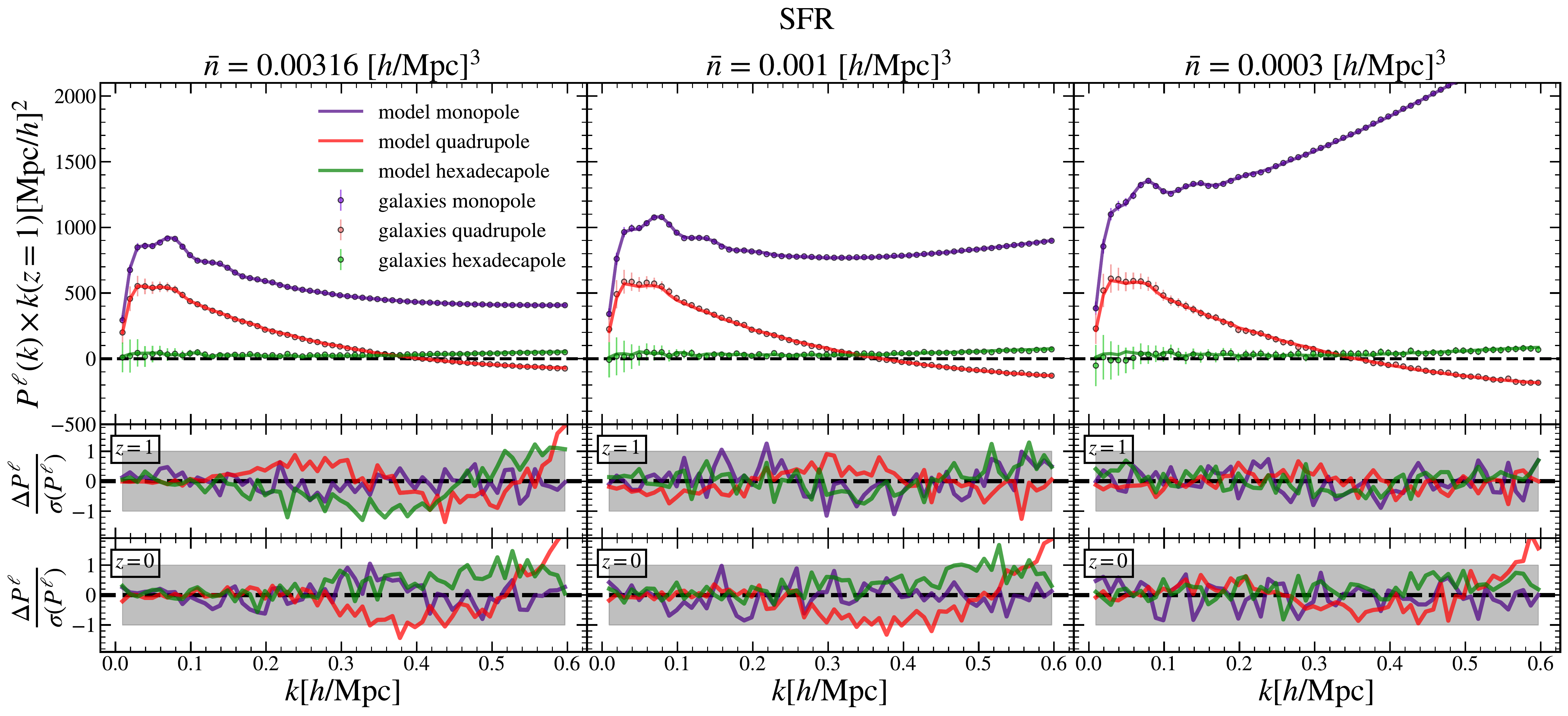}
	\includegraphics[width=\textwidth]{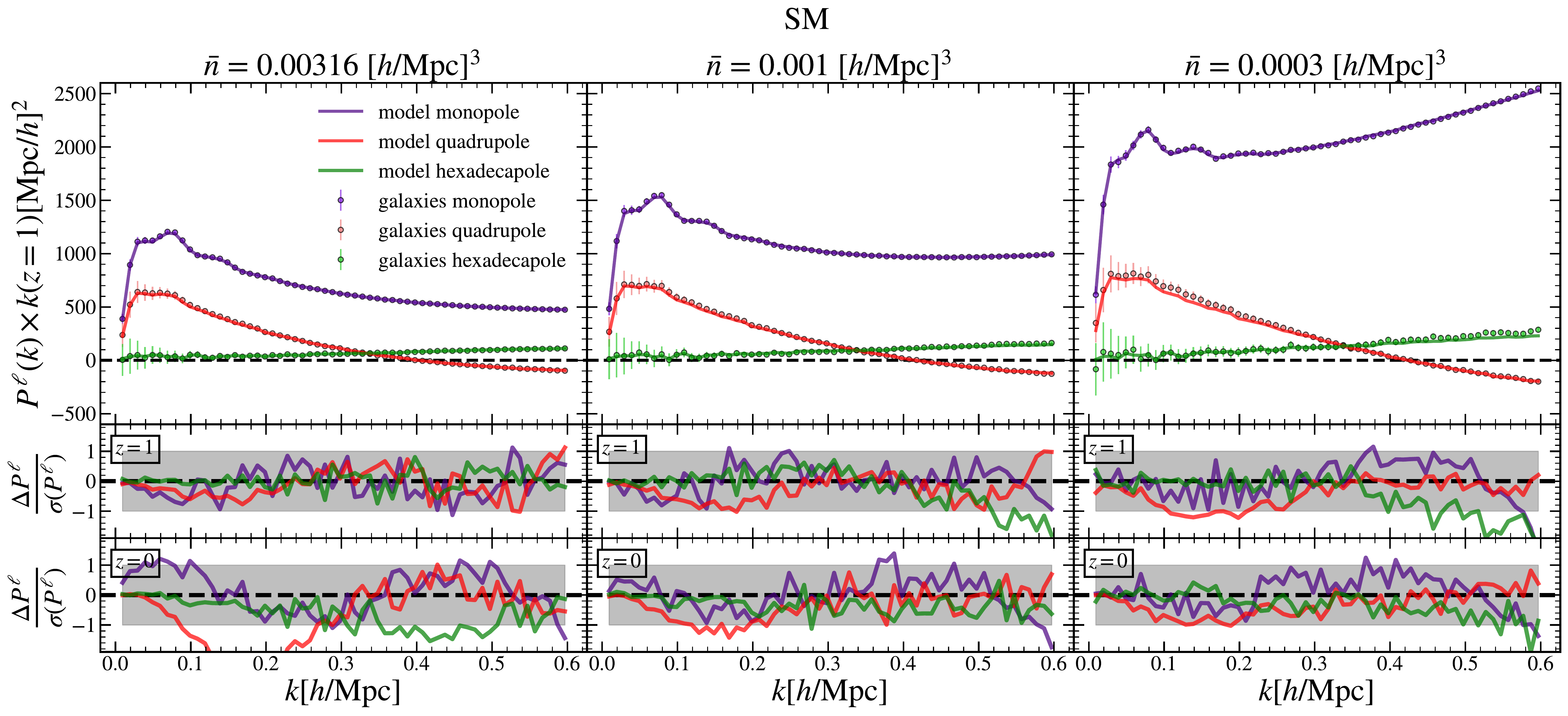}
    \caption{Fits of the model presented in this work to the monopole, quadrupole and hexadecapole of the power spectrum in redshift space for the Star Formation Rate and Stellar Mass selected galaxy-like samples at $z=0$ (lower sub-panels) and $z=1$ (higher sub-panels) at three number densities. Upper panels show the overall fit. Lower panels show the difference between model and data weighted by the diagonal terms of Eq.~\ref{eq:ps_cov}. We show in bigger panels the cases resembling current and near-future survey data (e.g. Euclid-like top plot and BOSS-like lower plot).}
    \label{fig:fits}
\end{figure*}

\begin{figure*}
	\includegraphics[width=\textwidth]{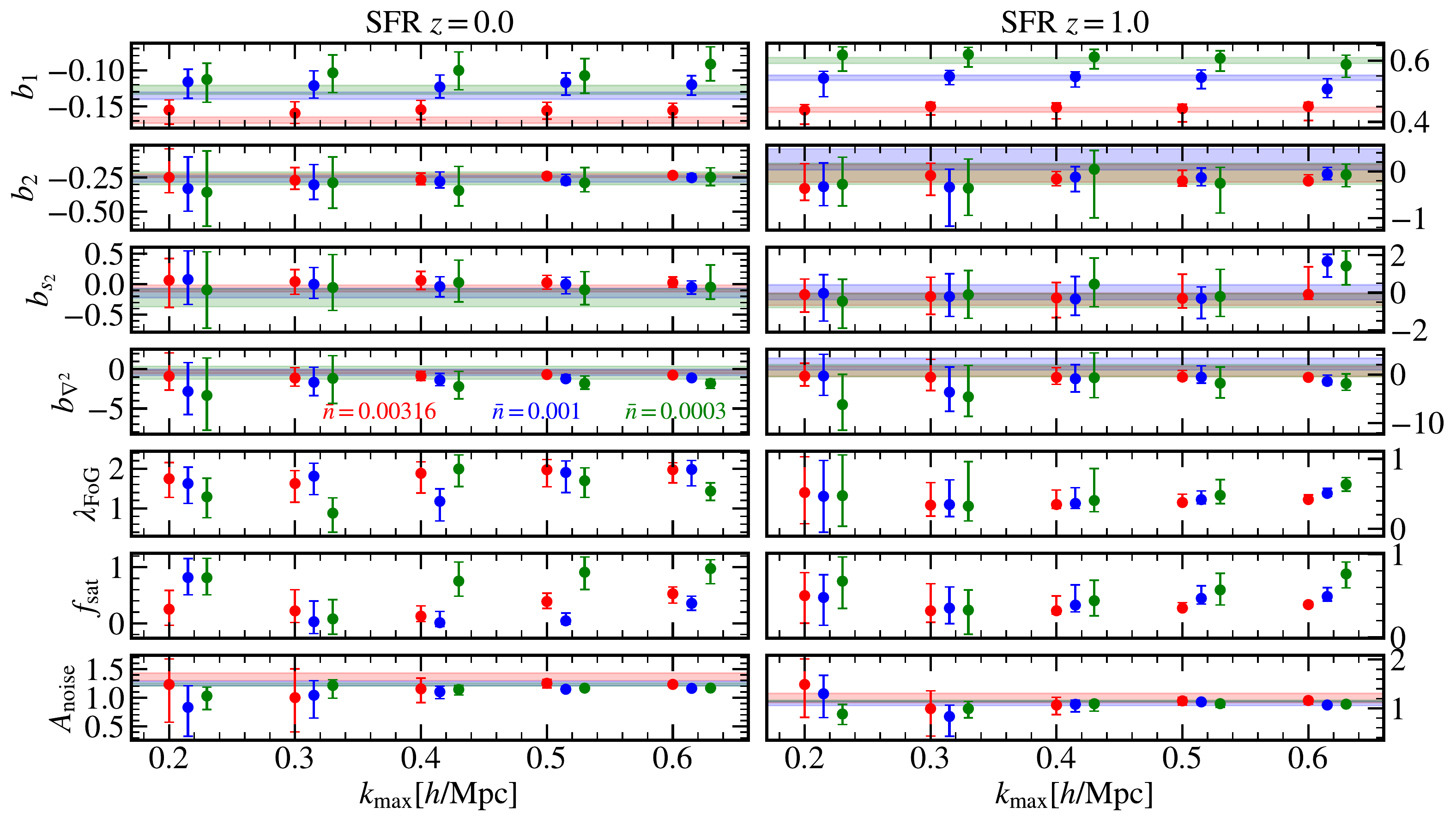}
	\includegraphics[width=\textwidth]{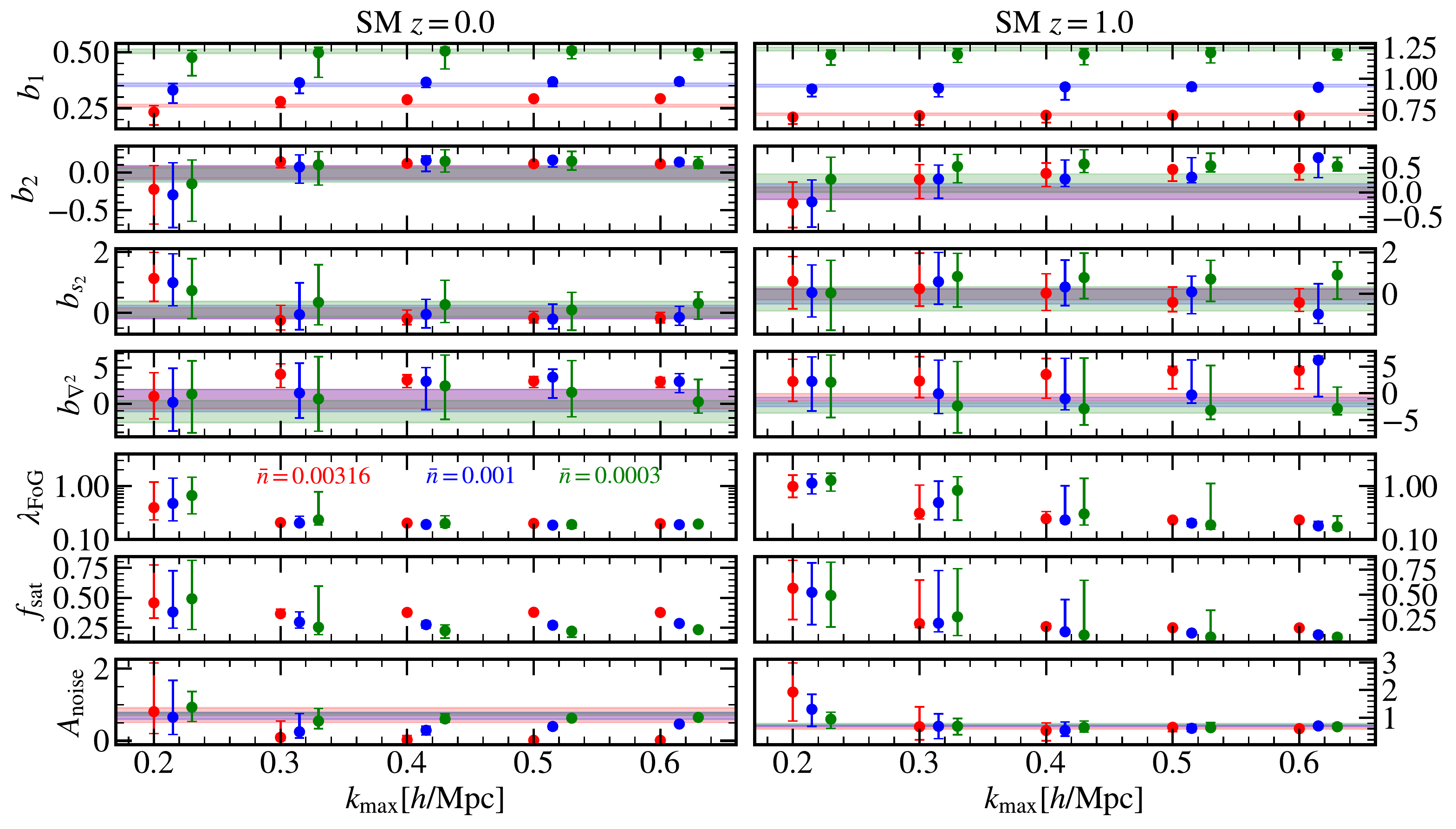}
    \caption{Fitted parameters at the number densities presented in the legend for SFR and SM selected galaxies at $z=0$ and $z=1$. Points show the parameters measured in redshift space. Shaded regions show the same parameters as measured from real space at $k_{\rm{max}}=0.6$ (these are the values used for the tests in \S\ref{sec:realtoredshift}). The points have been slightly displaced to avoid cluttering. The values at which they are evaluated are $k_{\rm{max}}=[0.2,0.3,0.4,0.5,0.6]$ in units of [$h/$Mpc].}
    \label{fig:bias_vals}
\end{figure*}

\begin{figure*}
    \includegraphics[width=\textwidth]{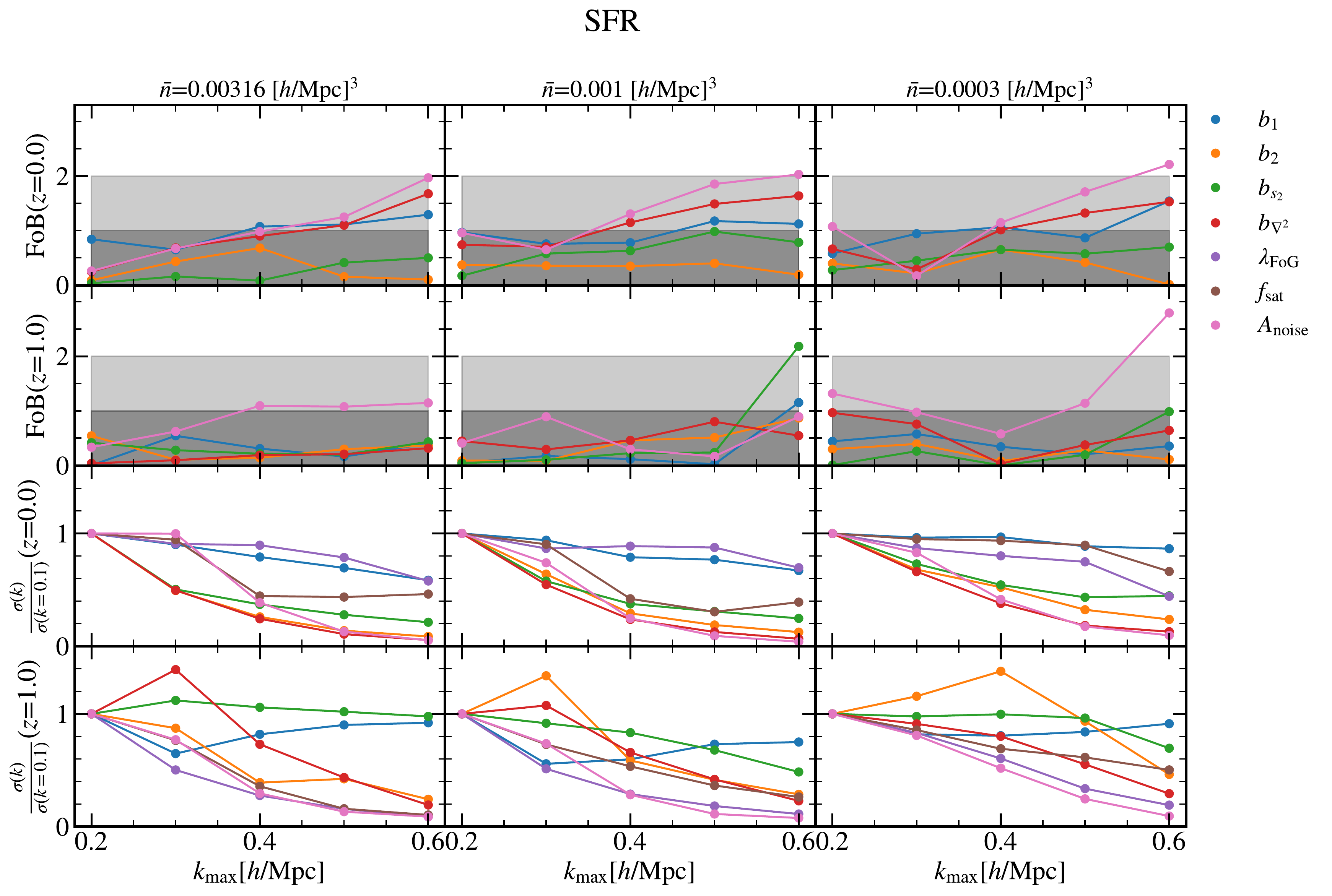}
    \caption{\textit{Upper panels}: Figure of Bias (FoB) at the number densities presented in the subtitles for the measured bias parameters in the SFR selected galaxies case at $z=0$ and $z=1$. Shaded regions show the 1-$\sigma$, 2-$\sigma$ regions. \textit{Lower panels}: Largest scale and smaller scales $\sigma$ comparison at the number densities presented in this work for SFR selected galaxies at $z=0$ and $z=1$.}
    \label{fig:FoBM_SFR}
\end{figure*}

We show how the model presented in \S\ref{sec:model} fits the SFR selected sample in the upper panels of Fig.~\ref{fig:fits}. At all number densities and redshifts explored here the model is able to reproduce the redshift space clustering at the two point level with remarkable accuracy all the way down to $k\approx 0.6 [h/$Mpc]. Moreover, we run \texttt{MULTINEST} over the likelihood defined in Eq~\ref{equation:clust_like} and show the results in the upper panels of Fig.~\ref{fig:bias_vals}. It shows in shaded regions the real space bias values and in dots the redshift space bias values. This Figure has several pieces of important information. 

We first point out the consistency of these parameters among different $k_{\rm{max}}$ values, never departing from the $1.5-\sigma$ region of the previous measurements. This shows that including smaller scales to this model does not statistically change the best-fitting values and no obvious tension between large and small scales appears.

Secondly, redshift space values roughly agree with their real space counterparts. Upper panels in Fig.~\ref{fig:FoBM_SFR} emphasise this point at both $z=0$ and $z=1$. These panels focus on the Figure of Bias (FoB) defined as

\begin{equation}
 {\rm{FoB}} = \frac{||b^{\rm{realspace}}_i(k_{\rm{max}}=0.6)|-|b^{\rm{z-space}}_i(k_{\rm{max}})||}{\sqrt{\sigma^2_{i,\rm{realspace}}(k_{\rm{max}}=0.6)+\sigma^2_{i,\rm{z-space}}(k_{\rm{max}})}}
 \label{eq:FoB}
\end{equation}

\noindent where $b_i$ belongs to $\{ b_1,b_2, b_s, b_{\nabla}, A_{\rm{noise}} \}$. The confidence regions $\sigma_i$ show the 68\% confidence levels. All parameters lie within 2-$\sigma$ regions of their real space counterparts (with the exception of $A_{\rm{noise}}$ at the smallest scales). This allows us to interpret the bias parameters in the same way as in real space since the redshift space parameters carry the weight of the mapping real-to-redshift space. 

Thirdly, roughly all errorbars shrink with the inclusion of smaller scales, leading to the conclusion that these extra scales include information helping to constrain bias parameters. This turns relevant because the better we can constrain bias parameter, the more we can hope to break degeneracies and ultimately constrain cosmological parameters. A steady increase in constraining power is found for all parameters. The increase depends on the number density, finding a greater improvement in the denser number densities and lower redshift. To further assess the small scales constraining power on the bias parameters we show in lower panels of Fig.~\ref{fig:FoBM_SFR} the $\sigma$'s ratio as defined by $\sigma(k)/\sigma(k=0.1[h/\rm{Mpc}])$. As an example, if this value reduces to 0.5 , it would mean that the constraining power of smaller scales reduces the uncertainty by a half. Another puzzling feature is the increase of $\sigma$ in some few cases, such as $\sigma_{b_2}$ at $k=0.3[h/$Mpc] in $\bar{n}=0.001[h/\rm{Mpc}]^3$. We associate this behaviour with noise in our chains due to unconstrained parameters and non-symmetric posteriors. We tested it by running different \texttt{MULTINEST} runs and found shifts in these values of less that 10\%. We further tested this by removing the unconstrained parameters from the fitting procedure and found an agreement with a slow but steady increase in constraining power. We decided to show here the results of the full model since it is needed to describe all cases. 

The results of this section suggest that our approach provides an accurate modelling of Emission Line Galaxies at $z\le 1$ and $k\approx 0.6 [h/$Mpc], and therefore is well suited to analyse the galaxy clustering of near-future surveys such as Euclid and DESI.

\subsubsection{Stellar Mass selected galaxies results}

Here we move to SM selected galaxies and show how the model presented in \S\ref{sec:model} fits them in the lower plot of Fig.~\ref{fig:fits}. Note first that roughly all fits reproduce the redshift space clustering of the galaxy sample within the errorbars. Focusing on $z=0$ and $\bar{n}=0.00316[h/\rm{Mpc}]^3$, this fit agrees better with the data than those of the left panel of Fig.~\ref{fig:Validation_SM_z0_shuffletests}, indicating absorption of extra dependencies coming from redshift space in the real space defined bias parameters. This suggests that the meaning of these parameters in terms of their real space counterparts is no longer valid at this number density. It also shows that the bias model has the ability to absorb redshift space dependencies that have not been taken into account. 

In order to explore which parameter is absorbing these dependencies, we show the results of the measured parameters in lower plots of Fig.~\ref{fig:bias_vals}. Redshift space values differ from their real space counterparts in several cases. To make this point clear we show in Fig.~\ref{fig:FoBM_SM} of Appendix \ref{App:SM_biases} the Figure of Bias defined in Eq.~\ref{eq:FoB}. Most parameters lie within 2$\sigma$ regions, but $A_{\rm{noise}}$ departs from this region already before $k=0.3[h/\rm{Mpc}]$ at the highest number density. Moreover, the linear bias $b_1$ departs from this region already at $k=0.4[h/\rm{Mpc}]$. This suggests that both $A_{\rm{noise}}$ and $b_1$ are the free parameters absorbing most of the extra dependencies not accounted by the velocity model at the two-point level comparison.

Results improve with redshift, finding better fits at $z=1$ as seen from lower panels in Fig.~\ref{fig:fits}. Moreover, from the right panels of the lower plot in Fig.~\ref{fig:bias_vals} and lower panels from Fig.~\ref{fig:FoBM_SM}, there is an agreement within $2-\sigma$ of roughly all parameters at all $k_{\rm{max}}$ as compared to their real space counterparts. Therefore, the model is able to reproduce SM clustering statistics all the way down to $k\approx 0.6 [h/$Mpc] at $z=1$. This might become important for surveys such as BOSS (\citealt{BOSS2017}).

Find also in the Appendix \ref{App:SM_biases} the $\sigma$'s ratio for these fits. An overall increase in constraining power on all parameters is found, showing the importance of including small scales in our analysis for LSS surveys. 

Note that all these results are sensitive to the errorbars assumed for the target clustering. The assumption of Gaussian errorbars on small scales is not accurate, thus this result cannot be taken as a prediction of the constraining power of the model. It is expected that small scales strongly correlate, making the errors at small scales larger and erasing constraining power. Nevertheless, no definitive answer can be known until a full study of covariance matrices is done all the way down to such small scales.
Strong efforts are currently being done on this direction (see e.g. \citealt{Balaguera2019b}, \citealt{Balaguera2019a}, \citealt{Pellejero2020a} and \citealt{Kitaura2020})

\section{Summary and Conclusions}
\label{sec:conclusions}

In this work we have proposed a new method for predicting the two-point statistics of galaxy catalogues in redshift  space as a function of the bias parameters $\{ b_1, b_2, b_s, b_{\nabla} \}$, redshift space specific parameters $\{ \lambda_{\rm{FoG}}, f_{\rm{sat}} \}$ and nuisance parameter $A_{\rm{noise}}$. In principle, this model can be applied to future large scale surveys to extract the cosmological information.

The method is based on the models presented in \cite{Modi_2020},  and that we further developed in \cite{ZennaroAnguloPellejero2021} (see also \citealt{Kokron2021}) which predict galaxy fields by treating the galaxy formation physics through flexible Lagrangian bias parameters and use the non-linear displacement fields from $N$-body simulations to advect these Lagrangian galaxy fields to Eulerian space. We have extended this model to redshift space by combining the information of DM and halo velocities from $N$-body simulations with a two parameter Finger of God effect, taking into account the satellite velocity dispersion $\lambda_{\rm FoG}$ and the satellite fraction $f_{\rm sat}$. 

We tested this model on two mock galaxy samples: one sample was set up to mimic stellar mass selected galaxies (thereby imitating a BOSS like survey) and another sample set up to mimic galaxies that are selected by star formation rate (thereby behaving similar to an Euclid-like survey), as presented in \cite{ContrerasAnguloZennaro2020}. Both selections were tested for three number densities $\{ 0.00316, 0.001, 0.0003 \}$ [$h/$Mpc]$^{3}$ and two redshifts $\{0, 1\}$. 

We found that the model is able to recover the two-point clustering of almost all studied cases within the $N$-body error-bars down to scales of $k\approx 0.6 [h/$Mpc].

Furthermore we have checked whether the bias model that would be inferred by fitting the clustering in real space as in \cite{ZennaroAnguloPellejero2021} is consistent with the redshift space ones. We have found a remarkable agreement  between the two approaches amongst many different fitting scales, showing that the real space parameters $\{ b_1, b_2, b_s, b_{\nabla} \}$ and the redshift space parameters $\{ \lambda_{\rm FoG} , f_{\rm sat} \}$ indeed account distinctly for real and redshift space effects. The only exception being Stellar Mass selected galaxies at $z=0$ for the denser samples. We track down this inconsistency to missing dependencies of Finger of God effects with halo properties in our model.

We conclude that the model would, allegedly, be able to reproduce the two-point clustering of samples such as Euclid or DESI all the way down to $k\approx 0.6 [h/$Mpc]. This is a remarkable result since previous analysis were stopped at $k\approx 0.2-0.3 [h/$Mpc] for the lack of precision in the theoretical model. This suggests that our model can be used to exploit the content of observational surveys to a much higher degree than has been possible so far.

However, to prove this last point, some further work is required: First, a full study of the bias parameters together with cosmological parameters is needed to show that the model is able to recover correct cosmologies. Since the model is based on an $N$-body simulation, it requires the use of hundreds or thousands of $N$-body simulations for its cosmological evaluation. This is beyond the capability of current computational power and shows the need of emulator approaches such as the ones presented in \cite{ZennaroAnguloPellejero2021} and \cite{Kokron2021} (see also similar approaches by \citealt{AricoLinearEmulator, AricoBaryonEmul}, and Contreras et al. in prep. or \citealt{Pellejero-Ibanez2020} for an alternative). Secondly, a more reliable covariance matrix is needed to fit small scales down to $k\approx 0.6 h/$Mpc. For example, promising approaches are those proposed by the \texttt{BAM} team in \cite{Balaguera2019a, Balaguera2019b}, \cite{Pellejero2020a}, \cite{Sinigaglia2020}, \cite{Sinigaglia2020}, and \cite{Kitaura2020}. Thirdly, there is the need to corroborate the precision and accuracy of the model on mocks of the selected surveys (see e.g. BOSS mocks \citealt{Kitaura2016}) and finally, its application to real data. 

\section*{Acknowledgements}
The authors acknowledge the support of the ERC-StG number
716151 (BACCO). MPI acknowledges the support of the ``Juan de
la Cierva Formaci\'on'' fellowship (FJC2019-040814-I). The authors also acknowledge the computer resources at MareNostrum and the technical support provided by Barcelona Supercomputing Center (RES-AECT-2019-2-0012 \& RES-AECT-2020-3-0014)

%%%%%%%%%%%%%%%%%%%%%%%%%%%%%%%%%%%%%%%%%%%%%%%%%%
\section*{Data Availability}

The inclusion of a Data Availability Statement is a requirement for articles published in MNRAS. Data Availability Statements provide a standardised format for readers to understand the availability of data underlying the research results described in the article. The statement may refer to original data generated in the course of the study or to third-party data analysed in the article. The statement should describe and provide means of access, where possible, by linking to the data or providing the required accession numbers for the relevant databases or DOIs.



\bibliographystyle{mnras}
\bibliography{rsd} % if your bibtex file is called example.bib

%\begin{thebibliography}{99}
%\end{thebibliography}

%%%%%%%%%%%%%%%%%%%%%%%%%%%%%%%%%%%%%%%%%%%%%%%%%%

%%%%%%%%%%%%%%%%% APPENDICES %%%%%%%%%%%%%%%%%%%%%

\appendix

\section{Stellar Mass selected galaxies results with shuffled velocities}
\label{App:SM_shuffle}

\begin{figure*}
    \includegraphics[width=\textwidth]{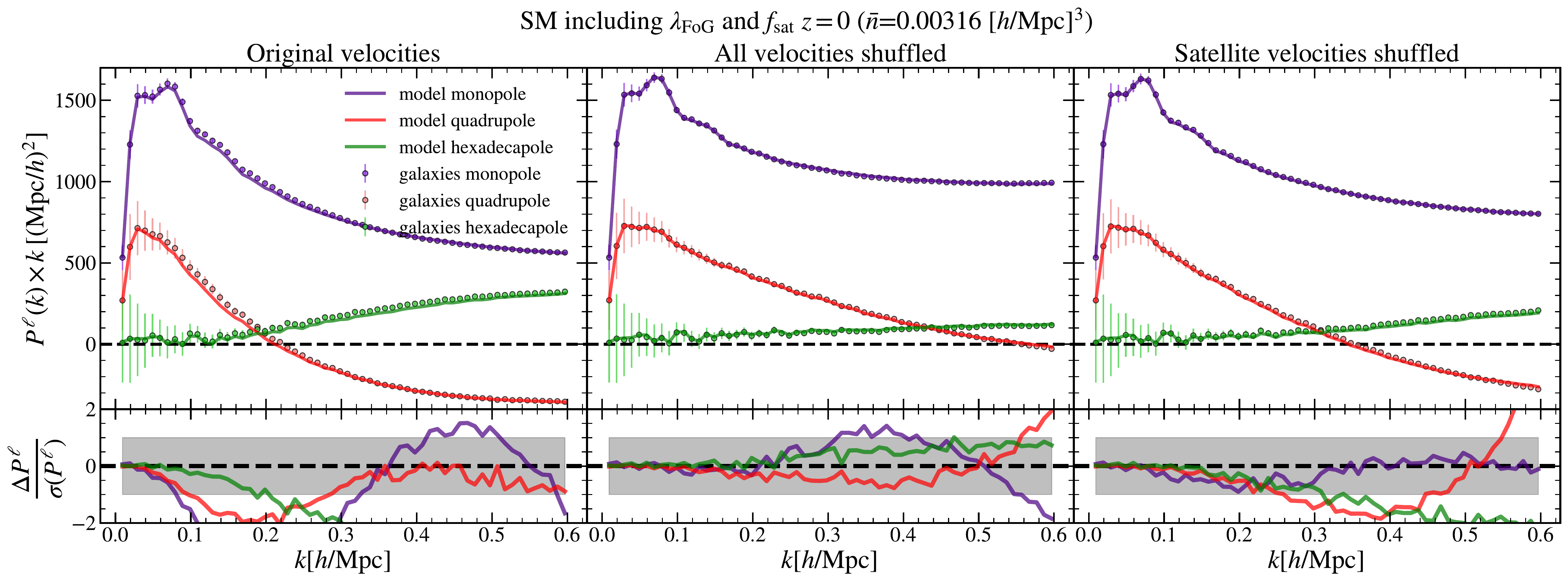}
    \caption{Results of fitting the model presented in \S\ref{sec:model} to the SM selected sample at $z=0$ and $\bar{n}=0.00316[h/\rm{Mpc}]^3$. Here we use the bias parameters measured from fitting power spectrum and cross-spectrum in real space $\{ b_1, b_2, b_s, b_{\nabla} \}$. Then $\lambda_{\rm{FoG}}$, $f_{\rm{sat}}$ and $A_{\rm{noise}}$ are included as free parameters. Left panel shows how the model performs on the original sample. Middle panel shows how this changes when shuffling all velocities of the galaxy sample. Right panel shows the same as middle panel but shuffling only the velocities among satellites.}
    \label{fig:Validation_SM_z0_shuffletests}
\end{figure*}

In this appendix we expand the results found in \S\ref{sec:realtoredshift} to the Stellar Mass selected sample. Left panel of Fig.~\ref{fig:Validation_SM_z0_shuffletests} shows that in this case the combined main halo-DM velocity of Eq.~\ref{eq:velocity} is not enough to reproduce the target clustering, given the real space measured bias parameters. We see clear departures from the target clustering. 

A key missing ingredient of the model presented in \S\ref{sec:model} is the dependence of the velocity dispersion on different properties of parent haloes. The clearest example of this is the dependence of the FoG effect on the mass of the host halo. In Eq.~\ref{eq:RSD}, $\lambda_{\rm{FoG}}$ is assumed to be free but the same for every object e.g. $\lambda_{\rm{FoG}} \neq \lambda_{\rm{FoG}}(M_{\rm{halo}}) $. To test this dependency and check if this is the origin of the discrepancy of our results for SM at $z=0$, we create two new samples based on the SM selected galaxies at $z=0$. Keeping the same positions as in the original catalogue but shuffling the relative velocities among objects, we create different catalogues in two ways. 

First, we randomly shuffle relative velocities among all objects independently of their mass and fit to the multipoles of the resulting sample. This mimics a catalogue where all velocities can be described by a simple constant FoG effect since any extra dependencies are removed by the shuffling. We show this result in the middle panel of Fig.~\ref{fig:Validation_SM_z0_shuffletests}. Here, including a $\lambda_{\rm{FoG}}$ makes the results able to reproduce the target galaxies clustering.

Secondly, we randomly shuffle the relative velocities among only satellites. This mimics a catalogue where the satellite velocities can be described by a simple constant FoG effect  since any extra dependencies are removed by the shuffling. We show this result in right most panel of Fig.~\ref{fig:Validation_SM_z0_shuffletests}. We find a better agreement between model and data in this case too although not as good as with the previous test. Some level of modelling might be missing in this step but a more thorough study might show that if we would have accounted for more realistic errors, the model would lie within them.

These two tests point to a lack of modelling of central mass halo properties such as mass as the main reason to the departures found in left-most panel of Fig.~\ref{fig:Validation_SM_z0_shuffletests}. A clear extension of this model would be to include the halo mass dependency into the $\lambda_{\rm{FoG}}$ parameter to account for this effect but we leave this for future work. Here we content ourselves with stating the level of inaccuracy of this approach and the extra error that should be included in the theoretical model for its use in the case of SM selected galaxies at $z=0$.

As a side note, and although not shown here, we tested this on the sparser number densities studied in this article. The effect decreases with number density and redshift. In the sparsest number density case ($\bar{n}=0.0003[h/\rm{Mpc}]^3$) including a $\lambda_{\rm{FoG}}$ and $f_{\rm{sat}}$ makes the results already indistinguishable from the target galaxies. This gives us hope to its application to real data SM selected galaxies such as the BOSS survey.

\section{Stellar Mass selected bias parameters}
\label{App:SM_biases}

\begin{figure*}
	\includegraphics[width=\textwidth]{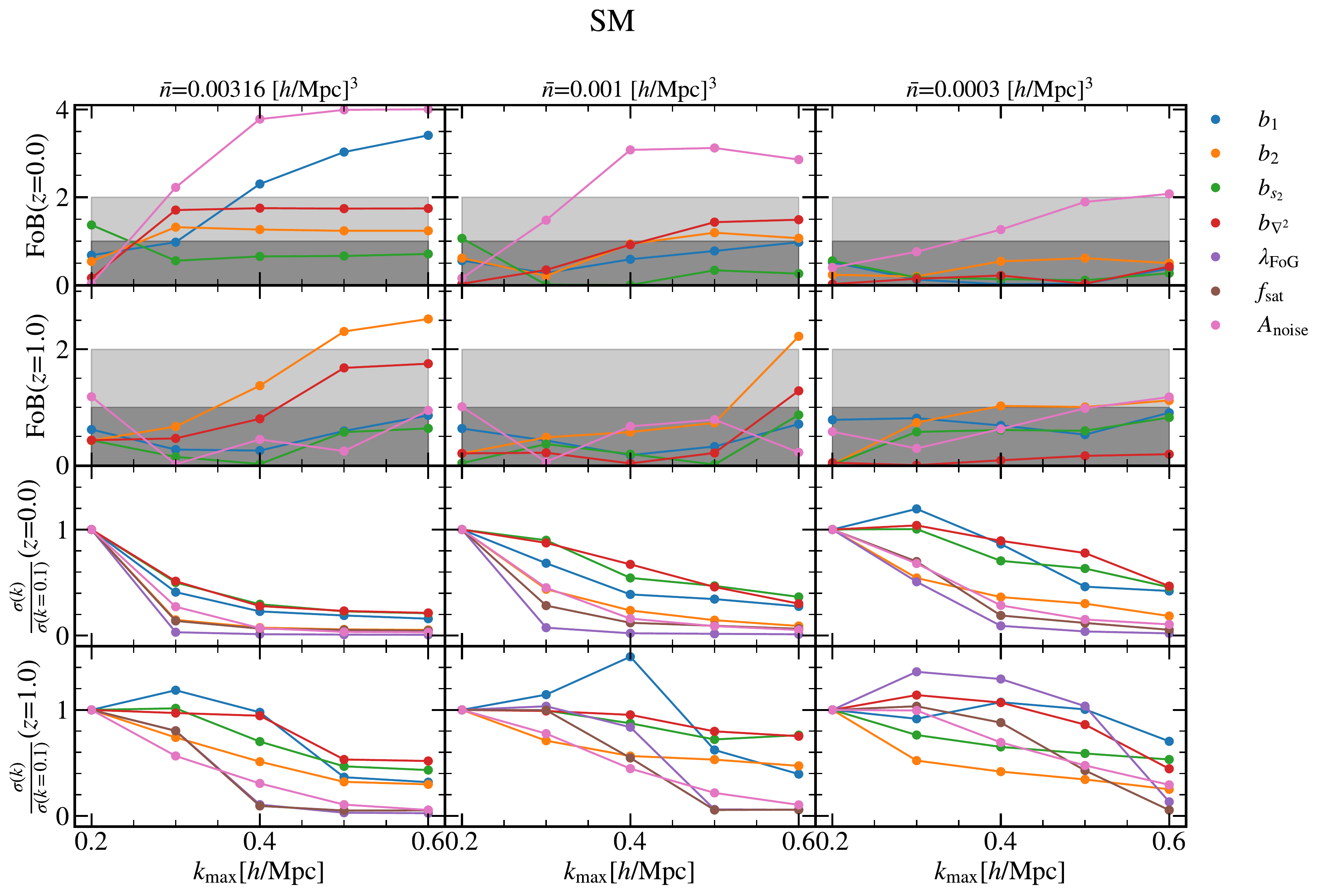}
    \caption{Same as Fig.~\ref{fig:FoBM_SFR} but for the Stellar Mass selected galaxies samples.}
    \label{fig:FoBM_SM}
\end{figure*}

In this appendix we expand the results found in \S\ref{sec:fitredshift} to the Stellar Mass selected galaxy sample. Find in Fig.~\ref{fig:FoBM_SM} a more concrete display of some of the features of the lower plot in Fig.~\ref{fig:bias_vals}. This Figure corresponds to Fig.~\ref{fig:FoBM_SFR} but for SM selected sample. In the upper panels of this figure we show the FoB as defined by Eq.~\ref{eq:FoB}. We find an overall larger departure of the redshift space measured parameters from their real space counterparts. This was somewhat expected, since the fits in redshift space reproduce the target clustering but that of the validation from real to redshift space does not. This then shows the level at which each parameter is affected by the lack of modelling. The clearest example is the behaviour of $b_1$ and $A_{\rm{noise}}$ at $\bar{n}=0.00316[h/\rm{Mpc}]^3$, which are in 4-$\sigma$ tension for some values of $k_{\rm{max}}$ although it is not the only example of greater than 1-$\sigma$ departure. The issue reduces with lower number densities, being still important at $\bar{n}=0.001[h/\rm{Mpc}]^3$ but disappearing at $\bar{n}=0.0003[h/\rm{Mpc}]^3$.

In the lower panels of Fig.~\ref{fig:FoBM_SM} find the ratio of the $\sigma$'s and hence the increase on constraining power factor depending on $k_{\rm{max}}$.

%%%%%%%%%%%%%%%%%%%%%%%%%%%%%%%%%%%%%%%%%%%%%%%%%%

% Don't change these lines
\bsp	% typesetting comment
\label{lastpage}
\end{document}